\documentclass[12pt]{article}
\usepackage{float,graphicx,amsmath,multirow}
\usepackage{color}
\usepackage[version=4]{mhchem}
\usepackage{booktabs}
\usepackage{times}
\usepackage{tablefootnote}
\usepackage{scicite}
\usepackage{microtype}
\usepackage{hyperref}
\graphicspath{{figures/}}

\newenvironment{sciabstract}{%
\begin{quote} \bf}
{\end{quote}}

\title{
    Interstellar Detection of the Highly Polar Five-Membered Ring Cyanocyclopentadiene }
\author{Michael C. McCarthy, Kin Long Kelvin Lee, Ryan A. Loomis, \\ Andrew M. Burkhardt,  Christopher N. Shingledecker, \\ Steven B. Charnley,
Martin A. Cordiner, Eric Herbst, \\ Sergei Kalenskii,  Eric R. Willis, Ci Xue,\\ Anthony J. Remijan, and Brett A. McGuire}

\begin{document}

\baselineskip24pt


\maketitle

\date{}

\begin{sciabstract}

Much like six-membered rings, five-membered rings are ubiquitous in organic chemistry, frequently serving as the building blocks for larger molecules, including many of biochemical importance.
From a combination of laboratory rotational spectroscopy and a sensitive spectral line survey in the radio band toward the starless cloud core TMC-1, we report the astronomical detection of 1-cyano-1,3-cyclopentadiene, $c$-\ce{C5H5CN}, a highly polar, cyano derivative of  cyclopentadiene, $c$-\ce{C5H6}. The derived abundance of $c$-\ce{C5H5CN} is far greater than predicted from astrochemical models which well reproduce the abundance of many carbon chains.  This finding implies either an important production mechanism or a large reservoir of aromatic material may need to be considered. The apparent absence of its closely-related isomer, 2-cyano-1,3-cyclopentadiene, may arise from its lower stability or be indicative of a more selective pathway for formation of the 1-cyano isomer, perhaps one starting from acyclic precursors.   The absence of N-heterocycles such as pyrrole and pyridine is discussed in light of the astronomical finding.

\end{sciabstract}

\section*{Main}

The recent astronomical detection of benzonitrile, $c$-\ce{C6H6CN} \cite{McGuire:2018it}, has at least partially resolved a long-standing conundrum in astrochemistry --- the apparent absence of five and six-membered rings, which are the building blocks of organic chemistry on earth.  For example, of the more than 135 M compounds registered in Chemical Abstract Service, it is estimated \cite{lipkus:4443,ruddigkeit:2864} that nearly 80\% contain at least one of these two rings.  In contrast, nearly 20 acyclic compounds with a comparable number of carbon atoms to benzene, $c$-\ce{C6H6}, were known astronomically~\cite{mcguire:census} prior to the discovery of benzonitrile.  Although benzene is nonpolar, and hence cannot be detected via its rotational transitions, its five-membered ring analog cyclopentadiene ($c$-\ce{C5H6}) is weakly polar, with a small permanent dipole moment ($\mu_b=0.420$\,D; Ref.~\citenum{laurie:635,damiani:265}), and a well-known rotational spectrum \cite{laurie:635}.  Nevertheless, like with benzene, replacing a single H atom in cyclopentadiene with a CN group imparts the three possible cyanocyclopentadiene variants with substantial dipole moments, regardless of the substitution site, meaning that their rotational lines are roughly 100 times more conspicuous relative to those of the hydrocarbon parent in space at the same abundance.

Unlike benzene, cyclopentadiene is a highly reactive diene which readily dimerizes at room temperature via a Diels-Alder reaction~\cite{turro:2841}.  It is widely used in synthetic organic chemistry, often for stereoselective purposes, while its anion $c$-\ce{C5H5^-}   is central to organometallic chemistry owing to the stability of metallocenes, such as derivatives of ferrocene \ce{Fe(C5H5)2}~\cite{werner:6052}, which are regularly used for the catalytic synthesis of asymmetric and chiral molecules.  Functionalized derivatives of cyclopentadiene have been the subject of many studies~[Ref.~\cite{dalkılıc:1966} and references therein]; substitution at the 1-carbon position is thermodynamically favored \cite{wentrup:4375} because it is fully conjugated with the $\pi$-electron system, as opposed to the other two positions which are either cross-conjugated (2-carbon) or not conjugated (5-carbon).


During the course of a laboratory study to identify products formed in an electrical discharge starting from benzene and molecular nitrogen~\cite{mccarthy:5170}, evidence was found for the two cyanocyclopentadienes shown in Fig.~\ref{figure:molecules}. Using cavity Fourier transform microwave spectroscopy, the rotational spectrum of each was subsequently measured at very high spectral resolution ($\sim$0.1\,ppm) between 5 and 40\,GHz, work which yielded highly accurate values for the three rotational constants, several of the leading quartic centrifugal distortion terms, and the two nitrogen quadrupole tensor elements (Tables~\ref{1-cpd_comparison} and \ref{2-cpd_comparison}).  Confirmation of the elemental composition and structure for both molecules is provided by the extremely close agreement of the rotational constants given in Table~\ref{table:constants} with those calculated theoretically and derived from previous measurements, although the earlier spectral analyses \cite{ford:326,sakaizumi:3903} were performed at considerably lower resolution (40-100\,kHz or 2-5\,ppm) compared to the present work (2\,kHz, or 0.1\,ppm).  With the new measurements, the rotational spectrum of both molecules can now be calculated to better than 0.05\,km~s$^{-1}$ at 20\,GHz in terms of equivalent radial velocity, more than sufficient for an astronomical search in the coldest, most quiescent molecular clouds where lines may be as narrow as 0.3--0.4\,km~s$^{-1}$ FWHM.

\begin{table}[t]
\caption{\label{table:constants}Comparison of experimental and theoretical rotational constants of 1- and 2-cyano-CPD. Constants are given in MHz, with values in parentheses corresponding to 1$\sigma$ uncertainty. The constants in this study were derived used a $A$-reduced Hamiltonian in the $I^r$ representation, A complete set of experimental spectroscopic constants and associated uncertainties is given in Table~\ref{1-cpd_comparison} and \ref{2-cpd_comparison}. Theoretical values are calculated at the $\omega$B97X-D/cc-pVQZ level of theory, with anharmonic vibrational corrections computed at the same level using second-order vibrational perturbation theory (VPT2).}
\begin{center}
\begin{tabular}{crrrr}
\toprule
& \multicolumn{4}{c}{1-cyano-CPD}\\
\cline{2-5}
Constant& \multicolumn{1}{c}{This Work}& \multicolumn{1}{c}{Ref.~\citenum{ford:326}}  & \multicolumn{1}{c}{Ref.~\citenum{sakaizumi:3903}} & \multicolumn{1}{c}{Theoretical} \\
\midrule
$A_0$  &   8352.98(2)     & 8356(5)     & 8353.97(9)& 8424.05  \\
$B_0$  &   1904.2514(3)   & 1904.24(2)  & 1904.24(1)& 1915.81   \\
$C_0$  &   1565.3659(3)   & 1565.36(2)  & 1565.38(1)& 1575.55   \\

\\
   & \multicolumn{3}{c}{2-cyano-CPD}\\ 
\cline {2-4}
 & \multicolumn{1}{c}{This Work} & \multicolumn{1}{c}{Ref.~\citenum{sakaizumi:3903}} & \multicolumn{1}{c}{Theoretical}\\
\midrule
$A_0$  &  8235.66(4) & 8235.0(13) & 8279.26 \\
$B_0$  &  1902.0718(2) & 1902.07(2) & 1916.03 \\
$C_0$ &  1559.6502(2) & 1559.67(2) & 1570.48 \\

\bottomrule
\end{tabular}
\end{center}
\end{table}

Concurrent with the laboratory work, a large-scale, high sensitivity spectral line survey, Green Bank Telescope Observations Hunting for Aromatic Molecules (GOTHAM), predominately in the K (18-27\,GHz) and K$_a$ bands (26-40\,GHz), has been underway since Fall 2018 toward the molecule-rich starless cloud core TMC-1 using the 100\,m Robert Byrd Green Bank Telescope (GBT) --- the same source and radio telescope used to detect benzonitrile \cite{McGuire:2018it}. Although only ${\sim}$30\% complete to date, where there is frequency coverage, it is roughly an order of magnitude more sensitive than a survey towards the same source using the Nobeyama 45\,m telescope \cite{Kaifu:2004tk}.  The higher sensitivity arises primarily from three factors: the larger collection area of the 100\,m dish, a beam size better matched to the source, and uniformly adopting a velocity resolution (0.05\,km~s$^{-1}$) that is appropriate in this narrow line source.  The resolution in the Nobeyama survey of 0.22--1.26\,km~s$^{-1}$ was frequently a factor of 2-4 times too low.  An overview of the GOTHAM survey, full observational details, and a discussion of the data reduction are described in \cite{McGuire:2020bb}.  

\section*{Results}

A detailed description of the data analysis procedures and statistics are presented in \cite{Loomis:2020aa} and summarized in the Methods. Briefly, a Markov Chain Monte Carlo (MCMC) fit is performed to the data using transitions of the target species which have predicted flux $\geq 5$\% of the strongest line.  In agreement with prior work \cite{Dobashi:2018kd,Dobashi:2019ev}, we detect and fit four distinct velocity components in the source (after accounting for hyperfine structure) with four column densities and four source sizes.  A uniform linewidth and excitation temperature is used to reduce the number of parameters being fit.  The MCMC fit to 1-cyano-CPD showed a significant detection in in all four velocity components.  This is shown visually in Fig.~\ref{1-cyano-CPD_stack}, where we see compelling evidence for 1-cyano-CPD, where as the isomer 2-cyano-1,3-cyclopentadiene (2-cyano-CPD) is not strongly detected.  

At our current detection sensitivity, we estimate an upper limit for 2-cyano-CPD roughly 1/3 that of 1-cyano-CPD, taking into account relatively small differences in the dipole moments and partition functions between the two. As such, it is not yet possible to establish with certainty if 1-cyano-CPD is formed selectively or if the apparent absence of  2-cyano-CPD is simply a question of sensitivity.  Although it may have no bearing on the astronomical observations, 2-cyano-CPD was found to be approximately four times less abundant than the 1-cyano-CPD in our laboratory study~\cite{mccarthy:5170} under identical experimental conditions; in the earlier study by Sakaizumi \textit{et al.} \cite{sakaizumi:3903}, the ratio was 2:1, with 1-cyano-CPD also being more abundant. 

\begin{figure}
     \centering
    \includegraphics[width=\textwidth]{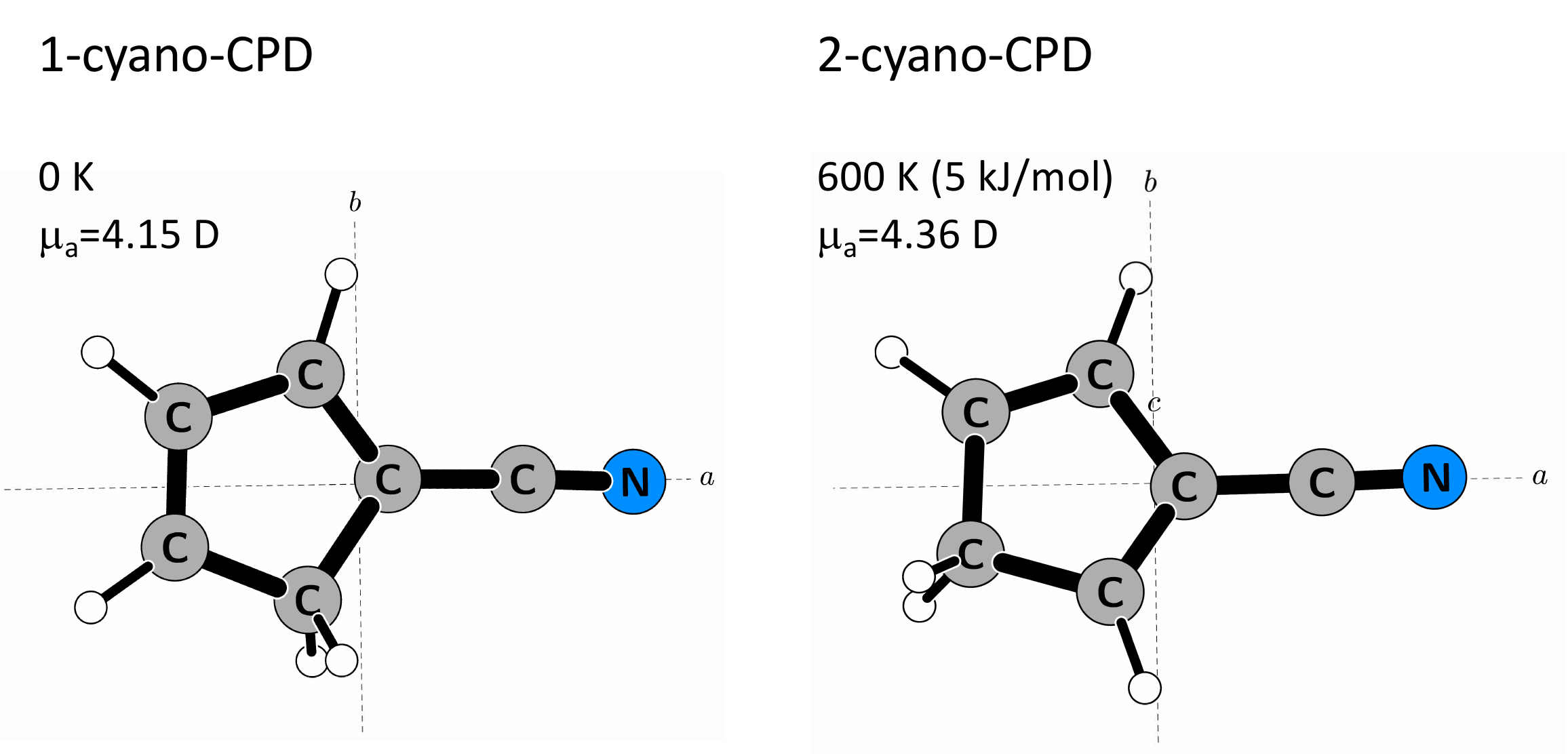}
     \caption{Geometric structures of the two low-lying cyanocyclopentadiene isomers, along with their relative stabilities calculated theoretically at the G3//B3LYP level of theory.  Owing to its substantially lower stability, the third isomer, 5-cyano-CPD, is not shown. Dipole moments are from Ref.~\citenum{sakaizumi:3903}.  }
     \label{figure:molecules}
 \end{figure}

To quantify the relative stabilities of the  three cyanocyclopentadiene isomers, energetics have been calculated with the B3LYP variant of the G3 thermochemical method (G3//B3LYP), which has been shown to yield reliable and highly accurate (${\sim}$4\,kJ/mol) energetics \cite{simmie_benchmarking_2015,lee_gas-phase_2019}.  As indicated in Fig.~\ref{figure:molecules}, 1-cyano-CPD is the most stable of the three, with 2-cyano-CPD computed to lie higher in energy by only 5\,kJ mol$^{-1}$ (600\,K), which is consistent with the energy ordering from the NMR findings of Wentrup and Crow \cite{wentrup:4375}, and conjugation arguments.  5-cyano-1,3-cyclopentadiene (5-cyano-CPD) is considerably less stable  (26\,kJ/mol, or ${\sim}$3130\,K), and presumably explains why this isomer was not detected in our laboratory study or prior work; for this reason, it is not considered further.  Both 1- and 2-cyano-CPD have comparably large dipole moments along their $a$-inertial axis [4.15(15)\,D~and 4.36(25)\,D, respectively; Ref.~\citenum{sakaizumi:3903}].  

\section*{Astrochemical implications}

\begin{figure}
    \centering
  \includegraphics[width=\textwidth]{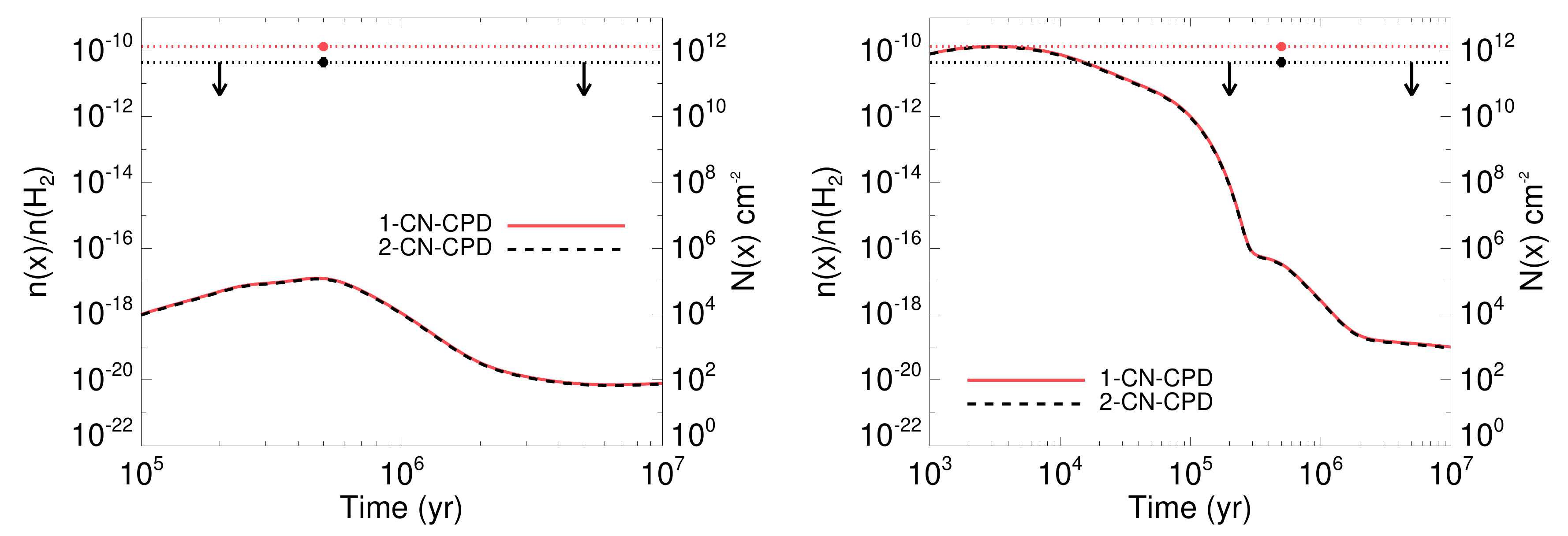}
    \caption{Abundance predictions for 1-cyano-CPD and 2-cyano-CPD from two chemical models in comparison to those derived from observations of TMC-1.  Results for 1-cyano-CPD are indicated in red, while those for 2-cyano-CPD are in black. \textit{Left panel}: model results from a full chemical treatment. Abundances derived from the MCMC analysis of the observational data are represented by dotted lines. \textit{Right panel}: Results from the same model but one including an initial abundance of CPD equal to $X(\ce{c-C5H6})_{t=0}=1.5\times10^{-7}$, or about 0.5\% of the total carbon budget, chosen to best reproduce the abundances obtained from our MCMC analysis.}
    \label{fig:modeling}
\end{figure}

Astrochemical modeling has been carried out to estimate the abundance of 1-cyano-CPD in TMC-1.  However, it is apparent from Fig.~\ref{fig:modeling} that the abundance inferred from observations vastly exceeds what our simulations predict, implying a far richer aromatic inventory for TMC-1 than previously expected. This result is surprising since the same models are in very good agreement with observations for many carbon-chain molecules at reasonable timescales for the source (10$^4$-10$^5$ years)\cite{Xue:2020aa,McGuire:2020bb}, with the long carbon chain \ce{HC11N} being a possible exception \cite{Loomis:2020aa} where the models actually overpredict the abundance relative to observation. The best agreement for 1-cyano-CPD is achieved at unrealistically early times in a simulation ($<10^4$ years) when a substantial initial abundance of CPD is assumed ($\sim$0.5\% of the carbon budget). Not surprisingly, the abundances of 1- and 2-cyano-CPD are nearly identical since, in the absence of a laboratory measurement, it was necessary to assume equal branching for cyanation of CPD. Even so, our models produce slightly more 1-cyano-CPD than 2-cyano-CPD due to the marginally larger dipole moment, and therefore faster rate of destruction with ions, of the latter isomer \cite{shingledecker_isomers_nodate}. 

$c$-\ce{C5H5CN} is now the fourth organic ring, along with benzonitrile and 1- and 2-cyanonaphthalene \cite{McGuire:2020aa}, that has been observed in TMC-1 with an abundance that exceeds, sometimes greatly so, predictions from our most up-to-date models.  Although not directly comparable, the same situation appears to be true for benzonitrile in other pre-stellar cores \cite{Burkhardt:2020aa}.  In light of these findings, other production pathways for cyclic molecules may need to be explored or a large unseen reservoir  of aromatic material, such as benzene, may need to be invoked. The surprisingly high abundances of cyclic species, deviations from previous astrochemical models, and promising potential pathways are further discussed in Ref. \citenum{Burkhardt:2020aa}.

In this context, it is intriguing that no evidence has been found in TMC--1 to date for well-known N-heterocycles (defined here as a cyclic compound in which one or more of the heavy atoms of the ring contains an atom other than carbon) of comparable size to 1-cyano-CPD and benzonitrile, such as pyrrole (\ce{C4H4NH}, $\mu=1.80$) and pyridine (\ce{C5H5N}, $\mu=2.19$; see Fig.~\ref{ulim_spec}),  differences in their polarity aside \cite{nbs}. Indeed,  upper limits of their column densities from the GOTHAM survey are of the same order as the measured column densities for 1-cyano-CPD and the cyanonaphthalene isomers \cite{McGuire:2020aa}. One possibility is that the lack of nitrogen incorporation into the heavy atom ring may indicate that much of the chemically active nitrogen is not available in the form of NH or \ce{NH2}, since the reaction of these radicals with butadiene \cite{lee_interstellar_2019} is thought to be an important pathway to pyrrole; an analogous argument may hold for pyridine.  A second possibility is that cynation of pre-existing aromatic material to produce $c$-\ce{C5H5CN} and other CN-functionalized rings is efficient.   Nevertheless, in either scenario, the preponderance of nitrile terminated chains and now rings highlights the key role CN radical apparently plays as a chemically active sink for nitrogen in TMC--1.



\newpage

\newpage

\section*{Acknowledgements}

The authors thank anonymous reviewers for helpful comments.  M.C.M. and K.L.K. Lee acknowledge support from NSF grant AST-1908576 and NASA grant 80NSSC18K0396. A.M.B. acknowledges support from the Smithsonian Institution as a Submillimeter Array (SMA) Fellow.   C.N.S. thanks the Alexander von Humboldt Stiftung/Foundation for their generous support, as well as V. Wakelam for use of the \texttt{NAUTILUS} v1.1 code.   S.B.C. and M.A.C. were supported by the NASA Astrobiology Institute through the Goddard Center for Astrobiology.  E.H. thanks the National Science Foundation for support through grant AST-1906489. C.X. is a Grote Reber Fellow, and support for this work was provided by the NSF through the Grote Reber Fellowship Program administered by Associated Universities, Inc./National Radio Astronomy Observatory and the Virginia Space Grant Consortium. Support for B.A.M. was provided by NASA through Hubble Fellowship grant \#HST-HF2-51396 awarded by the Space Telescope Science Institute, which is operated by the Association of Universities for Research in Astronomy, Inc., for NASA, under contract NAS5-26555. The National Radio Astronomy Observatory is a facility of the National Science Foundation operated under cooperative agreement by Associated Universities, Inc.  The Green Bank Observatory is a facility of the National Science Foundation operated under cooperative agreement by Associated Universities, Inc.

\clearpage


\section*{Contributions}

M.C.M. and K.L.K.L. performed the laboratory experiments, theoretical calculations, and wrote the paper with the help of A.M.B. and C.N.S.
A.M.B and B.A.M. performed the astronomical observations and subsequent analysis. E. H. determined and/or estimated rate coefficients and is the originator for many of the chemical simulations.
A.M.B. and C.N.S. contributed or undertook the astronomical modeling and simulations. E.R.W., M.A.C., S.B.C., S.K., C.X. and B.A.M. contributed to the design of the GOTHAM survey, and helped revise the manuscript.



\section*{Corresponding author and requests for materials}

Correspondence to M.C. McCarthy (mccarthy@cfa.harvard.edu) and B.A. McGuire (brettmc@mit.edu).

\newpage

\section*{Author Information}

\noindent\textbf{Center for Astrophysics $\mid$ Harvard~\&~Smithsonian, Cambridge, MA 02138, USA}\\
Michael C. McCarthy, Kin Long Kelvin Lee, Andrew M. Burkhardt,  Brett A. McGuire \\

\noindent\textbf{National Radio Astronomy Observatory, Charlottesville, VA 22903, USA}\\
Ryan A. Loomis, Anthony J. Remijan,  Brett A. McGuire \\

\noindent\textbf{Department of Physics and Astronomy, Benedictine College, Atchison, KS 66002, USA}\\
Christopher N. Shingledecker\\

\noindent\textbf{Center for Astrochemical Studies, Max Planck Institute for Extraterrestrial Physics, Garching, Germany}\\ 
Christopher N. Shingledecker\\

 \noindent\textbf{Institute for Theoretical Chemistry, University of Stuttgart, Stuttgart, Germany}\\
Christopher N. Shingledecker\\

\noindent\textbf{Astrochemistry Laboratory and the Goddard Center for Astrobiology, NASA Goddard Space Flight Center, Greenbelt, MD 20771, USA}\\
Steven B. Charnley and Martin A. Cordiner \\

\noindent\textbf{Institute for Astrophysics and Computational Sciences, The Catholic University of America, Washington, DC 20064, USA}\\
Martin A. Cordiner \\

\noindent\textbf{Department of Chemistry, University of Virginia, Charlottesville, VA 22904, USA}\\
Eric Herbst, Eric R. Willis, Ci Xue\\

\noindent\textbf{Department of Astronomy, University of Virginia, Charlottesville, VA 22904, USA}\\
Eric Herbst\\

\noindent\textbf{Astro Space Center, Lebedev Physical Institute, Russian Academy of Sciences, Moscow, Russia}\\
Sergei Kalenskii\\

\noindent\textbf{Department of Chemistry, Massachusetts Institute of Technology, Cambridge, MA 02139, USA}\\
 Brett A. McGuire \\


\clearpage

\begin{figure}
    \centering
    \includegraphics[width=\textwidth]{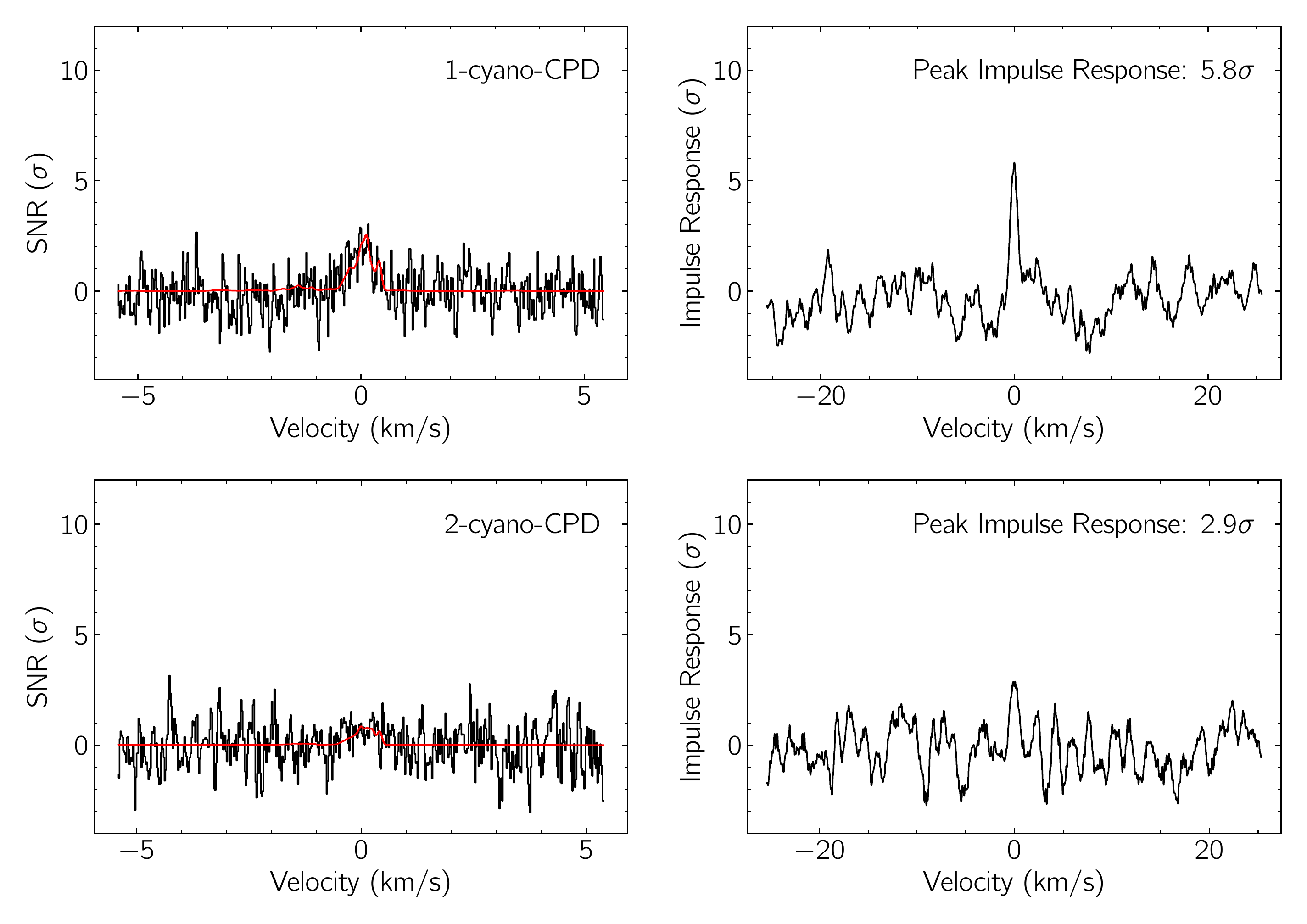}
    \caption{Velocity-stacked spectra of 1-cyano-CPD and 2-cyano-CPD and the impulse response function of the stacked spectra. \textit{Left panel}: Velocity-stacked spectra are in black, with the corresponding stack of the simulation using the best-fit parameters to the individual lines in red.  The data have been uniformly sampled to a resolution of 0.02\,km\,s$^{-1}$.  The intensity scale is the signal-to-noise ratio of the spectrum at any given velocity.  \textit{Right panel}: The impulse response function of the stacked spectrum generated using the simulated line profile as a matched filter.  The intensity scale is the signal-to-noise ratio of the response function when centered at a given velocity.  The peak of the impulse response function provides a minimum significance for the detection of 5.8$\sigma$. Bottom row shows the same, but for the 2-cyano-CPD upper limit (2.9$\sigma$). See Loomis et al.\cite{Loomis:2020aa} for details.}
    \label{1-cyano-CPD_stack}
\end{figure}

\clearpage
\newpage
\section*{Methods}

\subsection*{Spectral Stacking Routine}

Full details of the methodology used to detect and quantify molecules in our spectra are provided in Loomis et al. \cite{Loomis:2020aa}, including assumptions made in simulating the spectra and in determining physical parameters of the source, priors used in the fitting analysis, and an in-depth examination of the robustness of the methodology to interlopers and false-positives.  Briefly, we first perform a Markov-Chain Monte Carlo (MCMC) fit to the lines of strongly-detected and optically thin \ce{HC9N} and benzonitrile in our data.  Model spectra for these fiducial molecules, and for all other species, are generated using the formalisms outlined in Turner \cite{Turner:1991um}, which includes corrections for optical depth, and adjusted for the effects of beam dilution. The specific transition parameters for each species are obtained from spectral line catalogs primarily pulled from publicly accessible databases (\href{spec.jpl.nasa.gov}{spec.jpl.nasa.gov} and \href{https://cdms.astro.uni-koeln.de/classic/}{https://cdms.astro.uni-koeln.de/classic/}).  In some cases, we have generated these catalogs directly from the parameters provided from laboratory work outlined in individual publications.  In almost all cases, the substantial number of transitions used in the analysis (100s - 1000s) makes in impractical to provide a table of parameters in-text.  Instead, the interested reader is referred to the catalog files in the online supplementary data which contain all of the required information in a machine-readable format.

We detect four distinct velocity components ($v_{lsr}$) for nearly all molecules, in agreement with prior observations of the source \cite{Dobashi:2019ev}.  We also simultaneously fit for column densities ($N_T$), source sizes ($\theta_s$), excitation temperatures ($T_{ex}$), and linewidth ($\Delta V$).  The derived parameters for \ce{HC9N} and $c$-\ce{C6H5CN} are then used as Gaussian priors for MCMC fits to other linear and cyclic species of interest, respectively.  In general, we find that the linewidths ($\sim$0.1--0.3\,km\,s$^{-1}$) and excitation temperatures ($T_{ex} = $ 5--7\,K) are consistent across velocity components and molecular species.  Both this trend, and the derived values, are in excellent agreement with prior observations of the source \cite{Dobashi:2019ev,Gratier:2016fj,Kaifu:2004tk}.

For nearly all species, there are many more transitions covered by our spectra than are visible above the local RMS noise level of the observations.  We therefore extract a small portion of the observations centered around each spectral line, disregarding any windows that have a spectral feature $>$5$\sigma$, to avoid interloping signals from other species.  An SNR weighted average of these spectra is then performed based on the expected intensity of the line (derived from the MCMC parameters) and the local RMS noise of the observations.   For the purposes of this analysis, largely due to hyperfine splitting, we treat the signals on a per-line basis rather than a per-transition basis. This  result is that the stacked feature is somewhat broadened, as the hyperfine components and velocity components are not collapsed, but there is no over-counting of flux.   This results in a substantial increase to overall SNR, with the spectrum now encapsulating the total information content of all observed lines, rather than only that from the brightest lines.  Finally, the model spectra are stacked using identical weights, and that stacked model is used as a matched filter which is cross-correlated with the stacked observations.  The resulting impulse response spectrum provides a lower limit statistical significance to the detection.  Because the filter contains the same broadened hyperfine and velocity structure as the stack, there is no loss in significance.  Below, we provide a number of details pertinent to the MCMC fitting and stacking analyses performed for the specific molecules discussed in this work.

Table~\ref{app:lines} shows the total number of transitions (including hyperfine components) of the molecules analyzed in this paper that were covered by GOTHAM observations at the time of analysis and were above our predicted flux threshold of 5\%, as discussed in \cite{Loomis:2020aa}.  Also included are the number of transitions, if any, that were coincident with interfering transitions of other species, and the total number of lines used after excluding interlopers.  Observational data windowed around these transitions, spectroscopic properties of each transition, and the partition function used in the MCMC analysis are provided in the Harvard Dataverse repository \cite{GOTHAMDR1}.

\subsection*{Reaction pathways and energetics}

\begin{figure}
    \centering
    \includegraphics[scale=0.7]{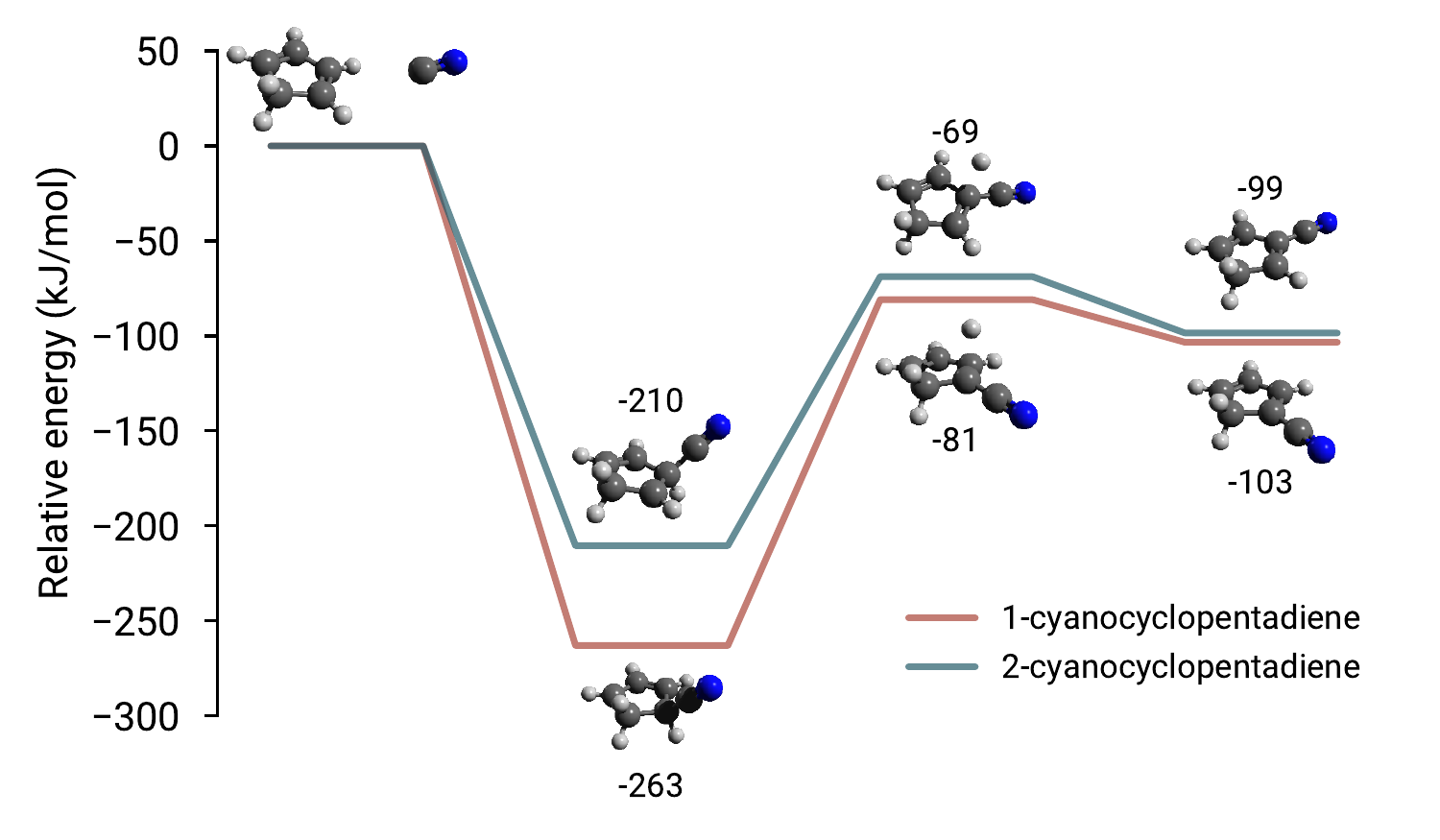}
    \caption{Thermochemistry (0\,K) for the reaction between \ce{CN} radical and CPD.  The calculated has been performed at the G3//B3LYP level of theory, and  energies in kJ/mol are given relative to the reactant asymptote. The reaction bifuricates as \ce{CN} attacks CPD barrierlessly, forming a radical intermediate. Subsequent hydrogen atom loss yields 1-cyano and 2-cyano-CPD.}
    \label{fig:cyano-pes}
\end{figure}

Possible formation pathways for 1-cyano- and 2-cyano-CPD under interstellar conditions have been investigated by calculating the reaction between cyclopentadiene and the CN radical, by analogy to benzonitrile formation~\cite{balucani:2000nw,woods:2002yx,Cooke:2020we}.  For this and other potential surface calculations, the growing string method \cite{zimmerman_growing_2013,zimmerman_automated_2013,jafari_reliable_2017} has been used: it is computationally efficient, allows various reactants to be rapidly surveyed and reaction barriers identified, and yields approximate reaction enthalpies between stationary points. Stationary points are then subsequently refined with the G3//B3LYP method. As shown in Fig.~\ref{fig:cyano-pes},  the $c$-\ce{C5H6} + \ce{CN} reaction to yield either 1-cyano or 2-cyano-CPD is found to be both exothermic and barrierless, irrespective of the stability of the the two isomers.

\begin{figure}
    \centering
    \includegraphics[scale=0.7]{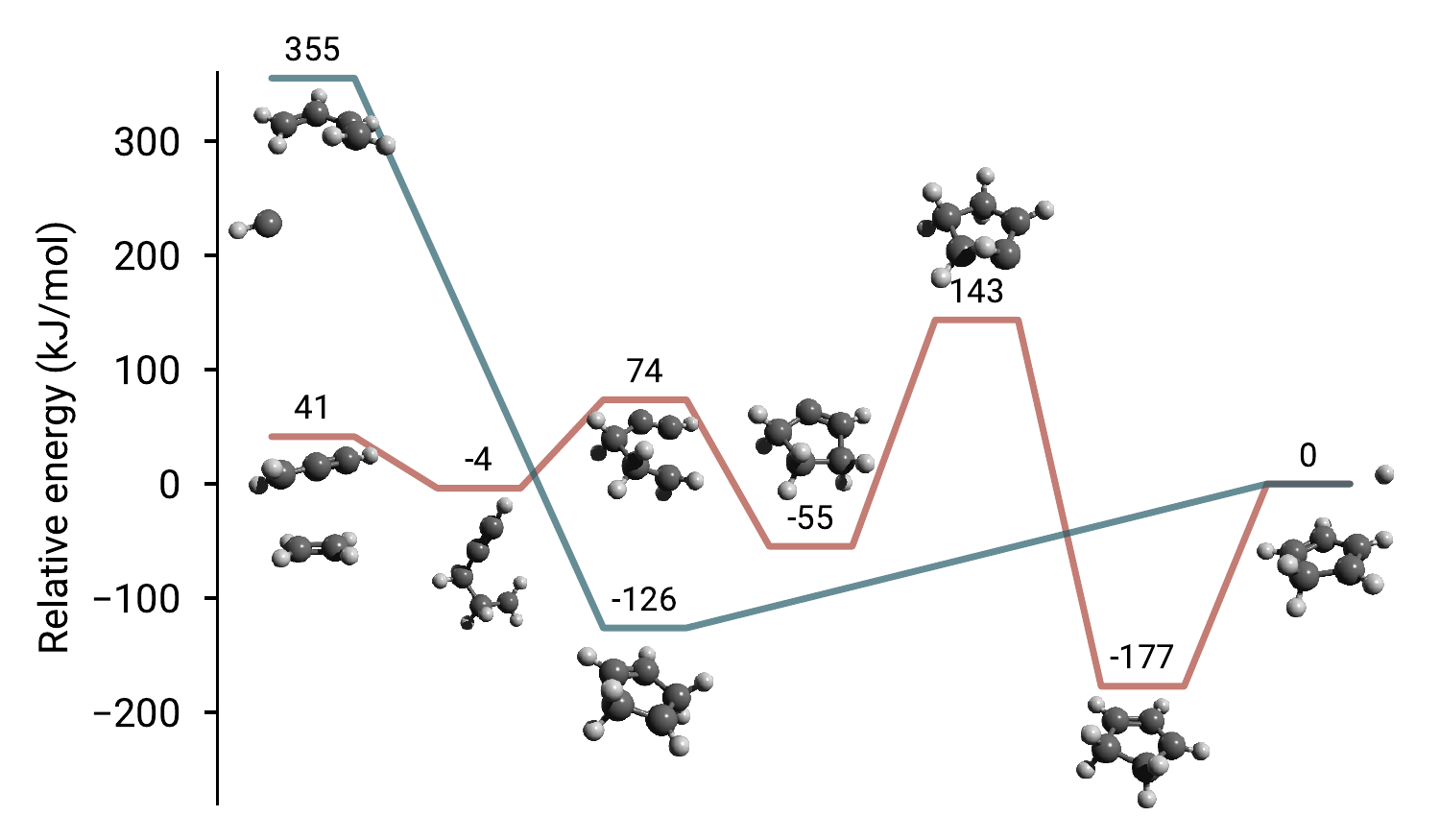}
    \caption{Potential energy surface for the formation of CPD at 0\,K. Reaction energies are given relative to the product (CPD + \ce{H} atom) asymptote. The red trace corresponds to reaction between propargyl radical (\ce{HCCCH2}) and ethylene (\ce{H2C=CH2}), and the blue trace represents \textit{gauche}-butadiene (\ce{H2C(CH)2CH2}) reacting with \ce{CH} radical.}
    \label{fig:cpd-formation}
\end{figure}

On the operative assumption that 1-cyano-CPD is formed from $c$-\ce{C5H6}, the reaction pathways that might produce this five-membered hydrocarbon ring in space have also been investigated, since they apparently have not been considered in current chemical models for dark clouds.  Two possible reactions are shown in  Fig.~\ref{fig:cpd-formation}. The red trace follows the reaction between ethylene (\ce{H2C=CH2}) and propargyl radical (\ce{HC#CCH2}), which combines barrierlessly to form an acyclic intermediate. Prior to CPD formation, however, this intermediate must undergo ring-closure which requires surmounting a barrier in excess of its initial energy (${\sim}$33\,kJ/mol, or 4000\,K), implying this pathway is not viable in TMC-1. The blue trace follows the reaction between \textit{gauche}-butadiene and \ce{CH} radical, which is calculated to produce CPD  exothermically and barrierlessly via subsequent hydrogen atom loss. Due to large energy barriers, the formation of CPD via reaction of allyl radical and acetylene \cite{bouwman_formation_2015} can also be neglected. It should be emphasized that reactions with large activation barrier may take place on grains by the bombardment of cosmic rays and internal UV photons \cite{shingledecker_cosmic-ray-driven_2018,shingledecker_general_2018}. Though we have not considered the effects of such reactions involving excited, suprathermal species here, it is possible that the contribution of these processes to the abundances of gas-phase aromatic molecules could be non-trivial, even in cold cores such as TMC-1.

\subsection*{Astrochemical Modeling}

We have used the \texttt{NAUTILUS}-v1.1 program \cite{ruaud_gas_2016}, together with a modified version the KIDA 2014 network \cite{wakelam_2014_2015} to predict the abundance of 1-cyano-CPD and other species in our GOTHAM observations \cite{McGuire:2020aa,Loomis:2020aa,Xue:2020aa}. 
This network includes the barrierless, exothermic reaction of CPD with CN radical to yield 1-cyano-CPD and 2-cyano-CPD , which we assume has a rate coefficient of $\sim3\times10^{-10}$ cm$^{-3}$ s$^{-1}$, which is reasonable assuming the reaction occurs with every collision \cite{Trevitt:2010ep,Cooke:2020we}.  Pathways that yield CPD from acyclic precursors are also explicitly included in our model, as are destruction reactions with ions \cite{woon_quantum_2009} and depletion onto grains.

Standard physical conditions relevant to TMC-1 were used in our simulations, e.g.,* a gas temperature of $T_\mathrm{gas}=T_\mathrm{dust}=10$ K, gas density of $2\times10^4$ cm$^{-3}$, and standard cosmic ray ionization rate of $1.3\times10^{-17}$ s$^{-1}$ \cite{Loomis:2016js, McGuire:2018it}. For initial elemental abundances, we use the values from Hincelin et al. \cite{hincelin_oxygen_2011}, except for the initial atomic oxygen, where we have chosen an initial value of $X(\mathrm{O})_{t=0}\approx1.5\times10^{-4}$, resulting in a slightly carbon rich C/O $\approx1.1$ by optimizing to best reproduce the observed cyanopolyyne abundances \cite{Loomis:2020aa}. To test the possible influence of CPD inherited from earlier stages of cloud evolution, we have also run a second simulation including an initial abundance of CPD, $X(\ce{c-C5H6})_{t=0}=1.5\times10^{-7}$, which was chosen to best reproduce the abundances obtained from our MCMC analysis. This is equivalent to approximately 0.5\% of the entire carbon budget in these simulations.



\section*{Data statement}
The datasets analyzed during the current study are available in the Green Bank Telescope archive (\href{https://archive.nrao.edu/archive/advquery.jsp}{https://archive.nrao.edu/archive/advquery.jsp}).  A user manual for their reduction and analysis is available as well (\href{https://greenbankobservatory.org/science/gbt-observers/visitor-facilities-policies/data-reduction-gbt-using-idl/}{https://greenbankobservatory.org/science/gbt-observers/visitor-facilities-policies/data-reduction-gbt-using-idl/}).  The complete, reduced survey data at X-band is available as Supplementary Information in \cite{McGuire:2020bb}.  The individual portions of reduced spectra used in the analysis of the individual species presented here is available in the Harvard Dataverse Archive \cite{GOTHAMDR1}.

\section*{Code statement}
All the codes used in the MCMC fitting and stacking analysis presented in this paper are open source and publicly available at \href{https://github.com/ryanaloomis/TMC1\_mcmc\_fitting}{https://github.com/ryanaloomis/TMC1\_mcmc\_fitting}.

\newpage
\clearpage
\section*{Extended Data}

\begin{table}[b]
    \centering
    \caption{Total number of transitions of a given species within the range of the GOTHAM data, number of interfering lines, and total number included in MCMC fit.}
    \begin{tabular}{l c c c}
    \toprule
                &   Transitions Covered &   Interfering Lines   &   Total Transitions   \\
    Molecule    &   By GOTHAM           &      In Data         &       Used in MCMC        \\
    \midrule
    1-Cyano-CPD &   111                &   0                   &   111    \\
    2-Cyano-CPD &   108                &   0                   &   108    \\
    cyclopentadiene &   21             &   0                   &   21     \\
    pyridine        &   76            &   0                   &   76    \\
    pyrrole         &   34             &   0                   &   34     \\
    \bottomrule     
    \end{tabular}
    \label{app:lines}
\end{table}

\newpage

\newpage
\clearpage

\part*{Supplementary Information}

\renewcommand{\figurename}{Supplementary Figure}
\renewcommand{\tablename}{Supplementary Table}
\setcounter{figure}{0}
\setcounter{table}{0}
\setcounter{equation}{0}

\section*{Analysis Results}
\label{app:}

The resulting best-fit parameters of the MCMC analysis carried out for 1-cyano-CPD are given in Supplementary Table~\ref{1-cyano-CPD_results}.  The stacked spectrum and matched filter results are shown in Supplementary Figure~\ref{1-cyano-CPD_stack}. A corner plot of the parameter covariances for the 1-cyano-CPD MCMC fit is shown in Supplementary Figure~\ref{1-cyano-CPD_triangle}.

\begin{table*}[bht!]
\centering
\caption{\ce{1-cyano-CPD} best-fit parameters from MCMC analysis}
\begin{tabular}{c c c c c c}
\toprule
\multirow{2}{*}{Component}&	$v_{lsr}$					&	Size					&	\multicolumn{1}{c}{$N_T^\dagger$}					&	$T_{ex}$							&	$\Delta V$		\\
			&	(km s$^{-1}$)				&	($^{\prime\prime}$)		&	\multicolumn{1}{c}{(10$^{11}$ cm$^{-2}$)}		&	(K)								&	(km s$^{-1}$)	\\
\midrule
\hspace{0.1em}\vspace{-0.5em}\\
C1	&	$5.602^{+0.007}_{-0.007}$	&	$120^{+69}_{-38}$	&	$4.09^{+0.89}_{-0.83}$	&	\multirow{6}{*}{$6.0^{+0.3}_{-0.3}$}	&	\multirow{6}{*}{$0.121^{+0.011}_{-0.011}$}\\
\hspace{0.1em}\vspace{-0.5em}\\
C2	&	$5.764^{+0.003}_{-0.003}$	&	$68^{+16}_{-15}$	&	$2.61^{+0.98}_{-0.94}$	&		&	\\
\hspace{0.1em}\vspace{-0.5em}\\
C3	&	$5.880^{+0.006}_{-0.006}$	&	$261^{+67}_{-81}$	&	$5.08^{+0.85}_{-0.82}$	&		&	\\
\hspace{0.1em}\vspace{-0.5em}\\
C4	&	$6.017^{+0.003}_{-0.003}$	&	$254^{+65}_{-83}$	&	$2.67^{+0.77}_{-0.78}$	&		&	\\
\hspace{0.1em}\vspace{-0.5em}\\
\midrule
$N_T$ (Total)$^{\dagger\dagger}$	&	 \multicolumn{5}{c}{$1.44^{+0.17}_{-0.17}\times 10^{12}$~cm$^{-2}$}\\
\bottomrule
\end{tabular}

\begin{minipage}{0.75\textwidth}
	\footnotesize
	\textbf{Note} -- The quoted uncertainties represent the 16$^{th}$ and 84$^{th}$ percentile ($1\sigma$ for a Gaussian distribution) uncertainties.\\
	$^\dagger$Column density values are highly covariant with the derived source sizes.  The marginalized uncertainties on the column densities are therefore dominated by the largely unconstrained nature of the source sizes, and not by the signal-to-noise of the observations.  See Fig.~\ref{1-cyano-CPD_triangle} for a covariance plot, and Loomis et al.\cite{Loomis:2020aa} for a detailed explanation of the methods used to constrain these quantities and derive the uncertainties.\\
	$^{\dagger\dagger}$Uncertainties derived by adding the uncertainties of the individual components in quadrature.
\end{minipage}

\label{1-cyano-CPD_results}

\end{table*}

\begin{figure*}
\centering
\includegraphics[width=\textwidth]{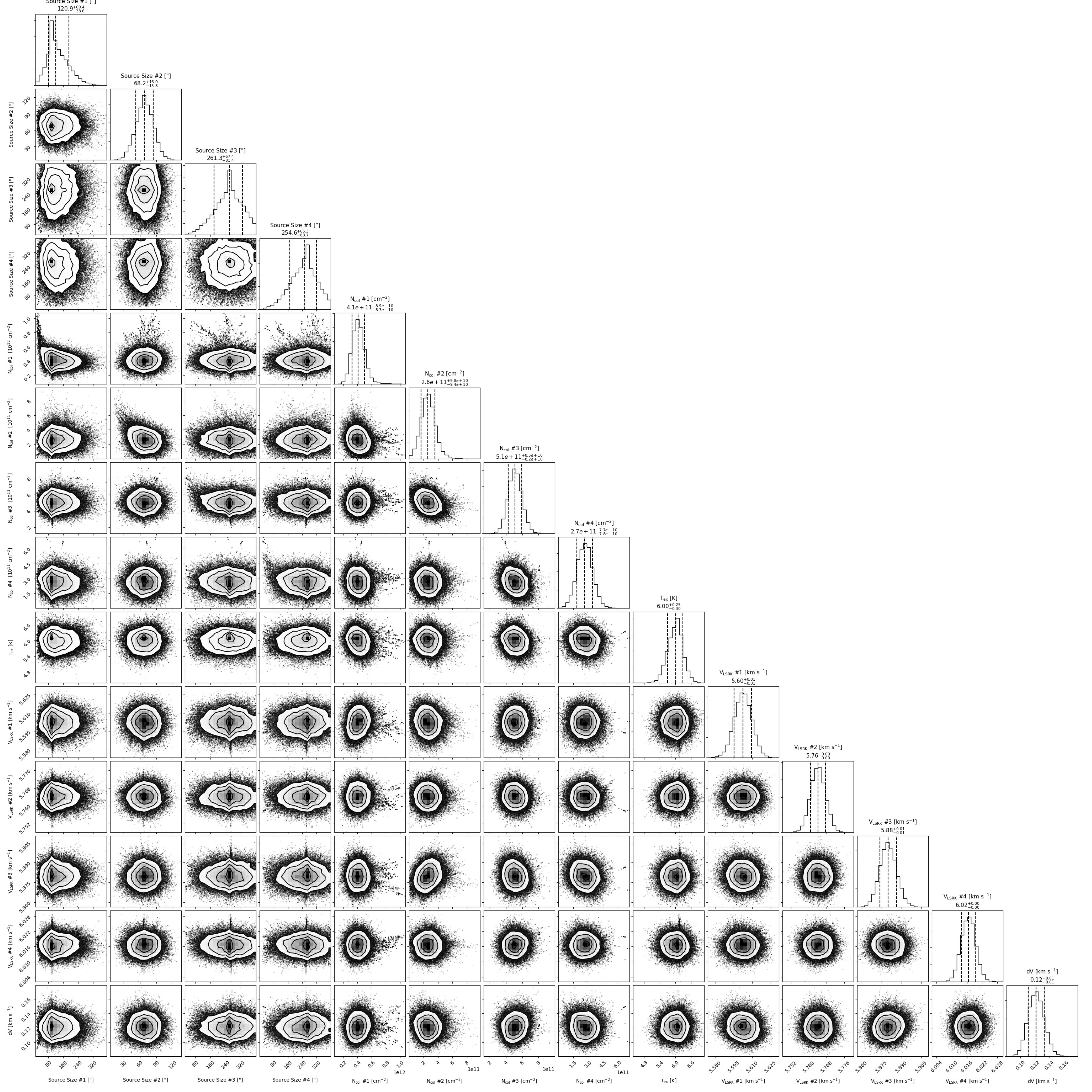}
\caption{Parameter covariances and marginalized posterior distributions for the 1-cyano-CPD MCMC fit. 16$^{th}$, 50$^{th}$, and 84$^{th}$ confidence intervals (corresponding to $\pm$1 sigma for a Gaussian posterior distribution) are shown as vertical lines. }
\label{1-cyano-CPD_triangle}
\end{figure*}

The resulting best-fit parameters of the MCMC analysis carried out for 2-cyano-CPD are given in Supplementary Table~\ref{2-cyano-CPD_results}.  The stacked spectrum and matched filter results are shown in Supplementary Figure~\ref{1-cyano-CPD_stack}. A corner plot of the parameter covariances for the 2-cyano-CPD MCMC fit is shown in Supplementary Figure~\ref{2-cyano-CPD_triangle}.

\begin{table*}
\centering
\caption{\ce{2-cyano-CPD} best-fit parameters from MCMC analysis}
\begin{tabular}{c c c c c c}
\toprule
\multirow{2}{*}{Component}&	$v_{lsr}$					&	Size					&	\multicolumn{1}{c}{$N_T^\dagger$}					&	$T_{ex}$							&	$\Delta V$		\\
			&	(km s$^{-1}$)				&	($^{\prime\prime}$)		&	\multicolumn{1}{c}{(10$^{11}$ cm$^{-2}$)}		&	(K)								&	(km s$^{-1}$)	\\
\midrule
\hspace{0.1em}\vspace{-0.5em}\\
C1	&	$5.593^{+0.008}_{-0.008}$	&	$101^{+72}_{-54}$	&	$<1.83^{+0.92}_{-0.70}$	&	\multirow{6}{*}{$6.1^{+0.3}_{-0.3}$}	&	\multirow{6}{*}{$0.128^{+0.012}_{-0.012}$}\\
\hspace{0.1em}\vspace{-0.5em}\\
C2	&	$5.764^{+0.003}_{-0.003}$	&	$66^{+16}_{-15}$	&	$<1.76^{+0.87}_{-0.75}$	&		&	\\
\hspace{0.1em}\vspace{-0.5em}\\
C3	&	$5.885^{+0.007}_{-0.007}$	&	$263^{+64}_{-85}$	&	$<0.77^{+0.60}_{-0.45}$	&		&	\\
\hspace{0.1em}\vspace{-0.5em}\\
C4	&	$6.017^{+0.003}_{-0.003}$	&	$251^{+66}_{-81}$	&	$<0.90^{+0.58}_{-0.51}$	&		&	\\
\hspace{0.1em}\vspace{-0.5em}\\
\midrule
$N_T$ (Total)$^{\dagger\dagger}$	&	 \multicolumn{5}{c}{$<5.26^{+1.51}_{-1.23}\times 10^{11}$~cm$^{-2}$}\\
\bottomrule
\end{tabular}

\begin{minipage}{0.75\textwidth}
	\footnotesize
	\textbf{Note} -- The quoted uncertainties represent the 16$^{th}$ and 84$^{th}$ percentile ($1\sigma$ for a Gaussian distribution) uncertainties.\\
	$^\dagger$Column density values are highly covariant with the derived source sizes.  The marginalized uncertainties on the column densities are therefore dominated by the largely unconstrained nature of the source sizes, and not by the signal-to-noise of the observations.  See Fig.~\ref{2-cyano-CPD_triangle} for a covariance plot, and Loomis et al.\cite{Loomis:2020aa} for a detailed explanation of the methods used to constrain these quantities and derive the uncertainties.\\
	$^{\dagger\dagger}$Uncertainties derived by adding the uncertainties of the individual components in quadrature.
\end{minipage}

\label{2-cyano-CPD_results}

\end{table*}

\begin{figure*}
\centering
\includegraphics[width=\textwidth]{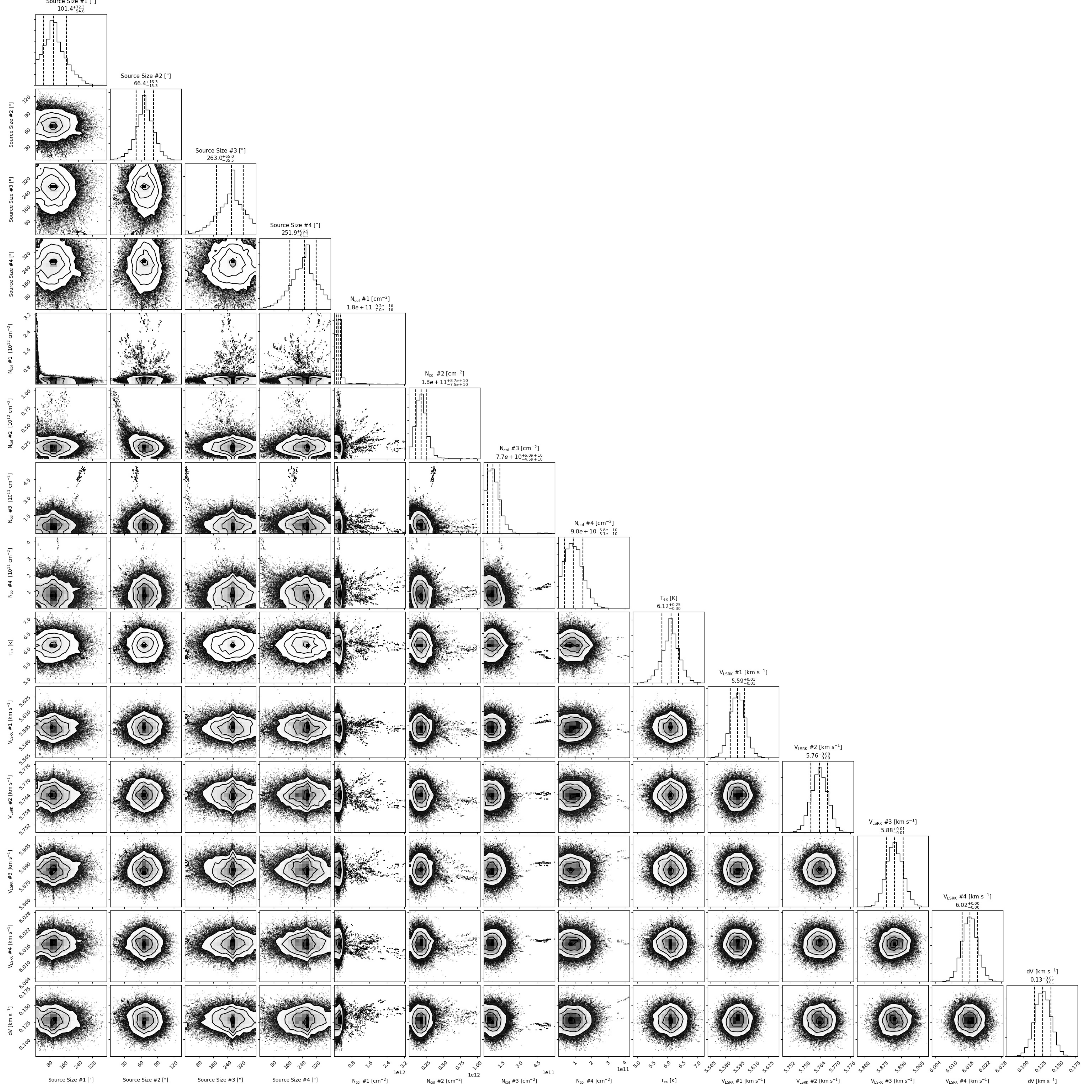}
\caption{Parameter covariances and marginalized posterior distributions for the 2-cyano-CPD MCMC fit. 16$^{th}$, 50$^{th}$, and 84$^{th}$ confidence intervals (corresponding to $\pm$1 sigma for a Gaussian posterior distribution) are shown as vertical lines. }
\label{2-cyano-CPD_triangle}
\end{figure*}

\clearpage

\section*{Upper Limits}

\begin{figure*}
    \centering
    \includegraphics[width=\textwidth]{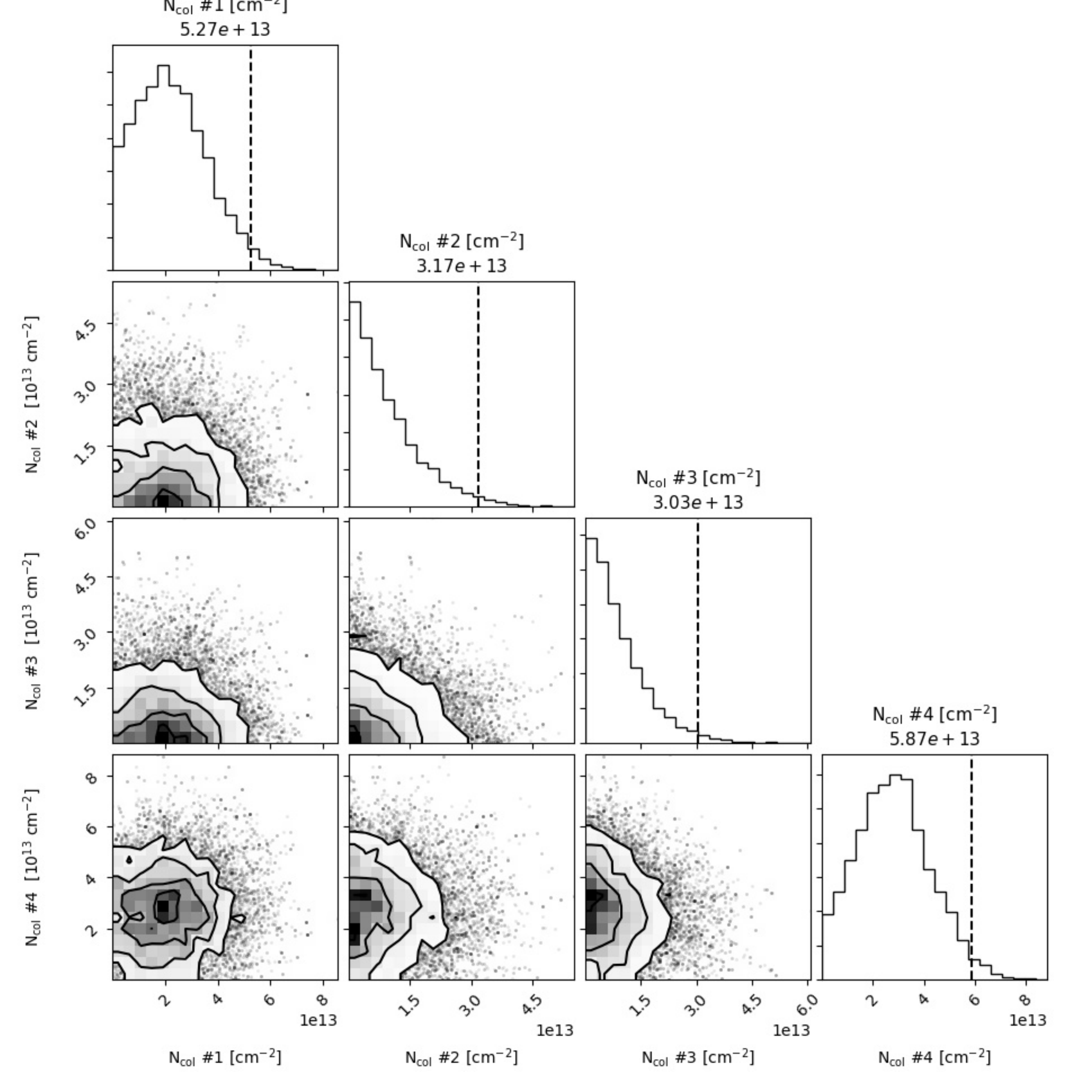}\\
    \caption{Corner plot of column densities in the four velocity components for cyclopentadiene.}
    \label{ulim_corner_cpd}
\end{figure*}

\begin{figure*}
    \centering
    \includegraphics[width=\textwidth]{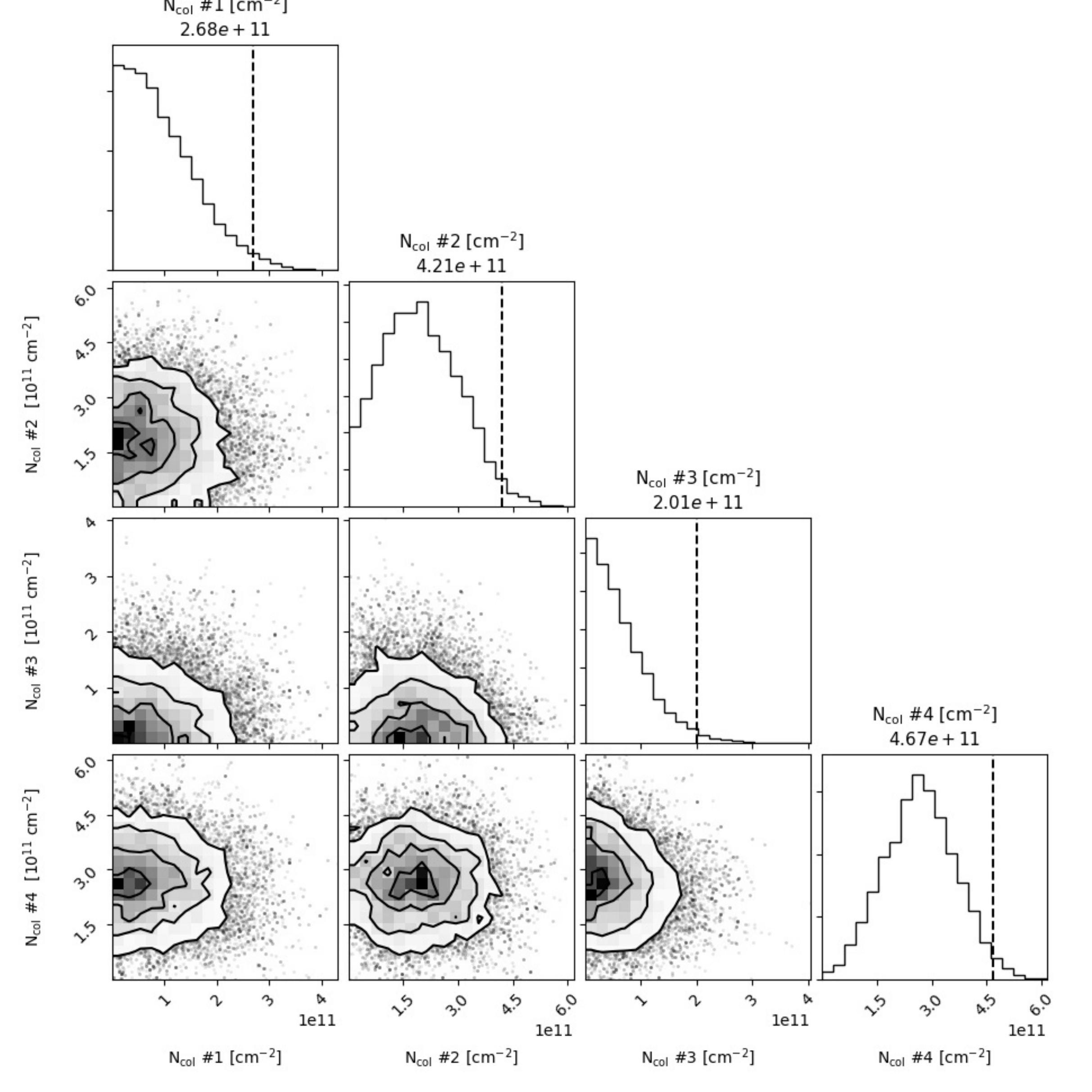}\\
    \caption{Corner plot of column densities in the four velocity components for pyridine.}
    \label{ulim_corner_pyri}
\end{figure*}

\begin{figure*}
    \centering
    \includegraphics[width=\textwidth]{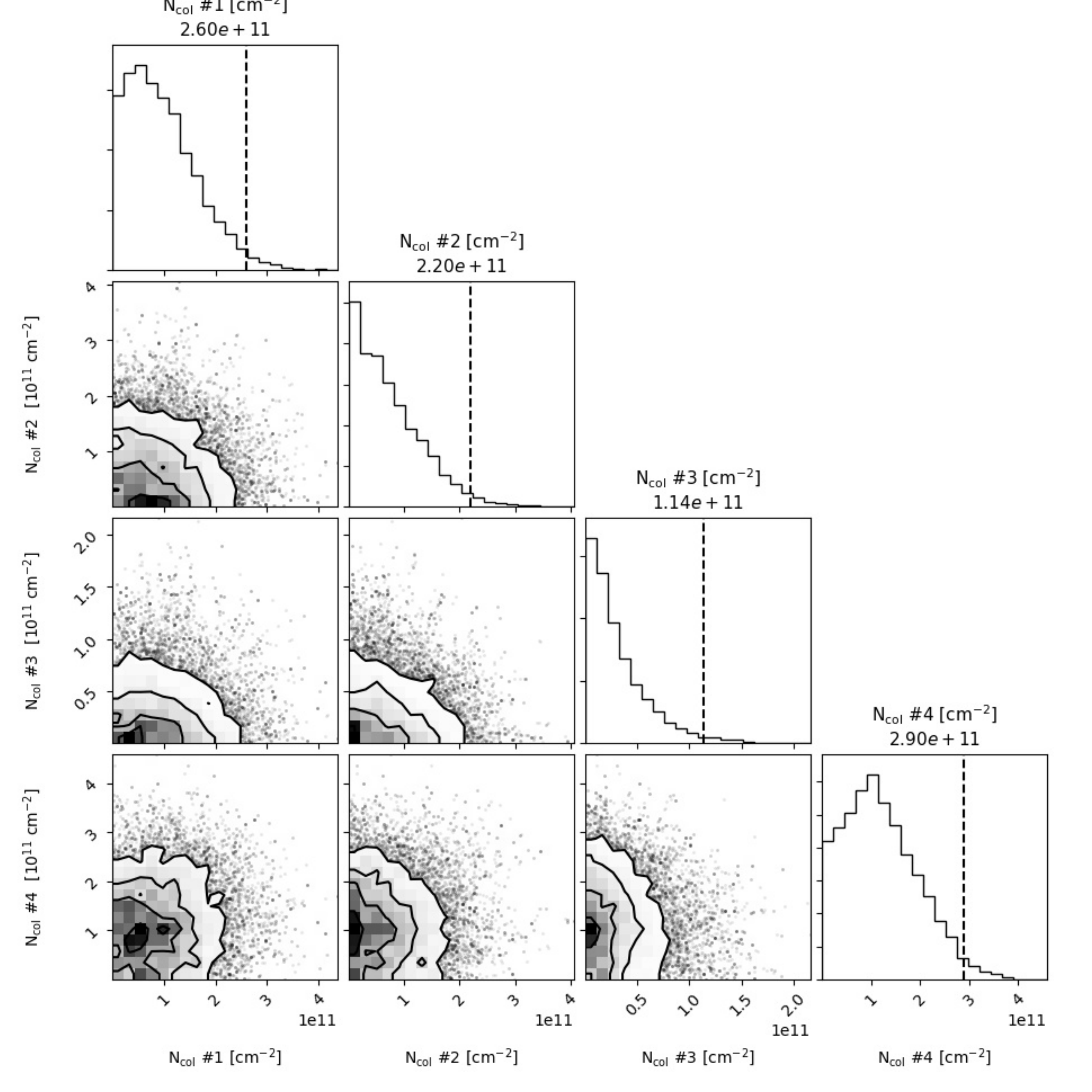}\\
    \caption{Corner plot of column densities in the four velocity components for pyrrole.}
    \label{ulim_corner_pyrr}
\end{figure*}

We have also calculated upper limits to the column densities of cyclopentadiene, pyridine, and pyrrole.  The stacked spectra, corner plots, and upper limits, including assumed parameters, are presented below in Supplementary Figure~\ref{ulim_spec}, Supplementary Figures~\ref{ulim_corner_cpd}-~\ref{ulim_corner_pyrr}, and Supplementary Table~\ref{ulim_table}, respectively.  While none of these show detectable signal in our current survey data, cyclopentadiene does display a small significance to a possible detectable column in one velocity component, as can be seen in its corner plot.  This may indicate that a detection may be possible with additional observation.

\begin{figure*}
    \centering
    \includegraphics[width=\textwidth]{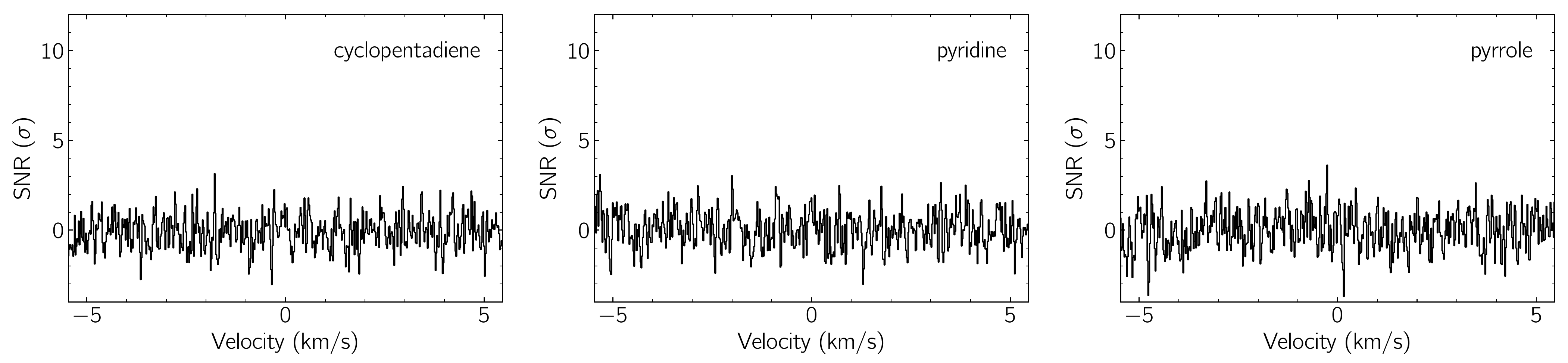}
    \caption{From left to right, velocity-stacked spectra of cyclopentadiene, pyridine, and pyrrole in black, using the best-fit parameters from the MCMC analysis for the upper limits.}
    \label{ulim_spec}
\end{figure*}


\begin{table*}
\centering
\caption{Cyclopentadiene, pyridine, and pyrrole best-fit parameters from MCMC analysis}
\begin{tabular}{c c c c c c c c}
\toprule
						&	&	&	&	&   cyclopentadiene	&	pyridine	&	pyrrole	\\
\multirow{2}{*}{Component}&	$v_{lsr}$					&	Size										&	$T_{ex}$							&	$\Delta V$		&	\multicolumn{1}{c}{$N_T^\dagger$}	&	\multicolumn{1}{c}{$N_T^\dagger$}	&	\multicolumn{1}{c}{$N_T^\dagger$} \\
			&	(km s$^{-1}$)				&	($^{\prime\prime}$)				&	(K)								&	(km s$^{-1}$)	&	\multicolumn{1}{c}{(10$^{11}$ cm$^{-2}$)}	&	\multicolumn{1}{c}{(10$^{11}$ cm$^{-2}$)}	&	\multicolumn{1}{c}{(10$^{11}$ cm$^{-2}$)} \\
\midrule
\hspace{0.1em}\vspace{-0.5em}\\
C1	&	[$5.595$]	&	[$92$]	&	[$6.1$]	&	[$0.121$]	&	$<$$214$	&	$<$$0.79$	&	$<$$0.95$\\
\hspace{0.1em}\vspace{-0.5em}\\
C2	&	[$5.764$]	&	[$68$]	&	[$6.1$]	&	[$0.121$]	&	$<$$70$		&	$<$$1.92$	&	$<$$0.56$\\
\hspace{0.1em}\vspace{-0.5em}\\
C3	&	[$5.885$]	&	[$265$]	&	[$6.1$]	&	[$0.121$]	&	$<$$71$		&	$<$$0.51$	&	$<$$0.24$\\
\hspace{0.1em}\vspace{-0.5em}\\
C4	&	[$6.016$]	&	[$250$]	&	[$6.1$]	&	[$0.121$]	&	$<$$276$	&	$<$$2.66$	&	$<$$1.16$\\
\hspace{0.1em}\vspace{-0.5em}\\
\midrule
\multicolumn{8}{c}{cyclopentadiene $N_T$ (Total)$^{\dagger\dagger}$: $630\times 10^{11}$~cm$^{-2}$}\\
\multicolumn{8}{c}{pyridine $N_T$ (Total)$^{\dagger\dagger}$: $5.9\times 10^{11}$~cm$^{-2}$}\\
\multicolumn{8}{c}{pyrrole $N_T$ (Total)$^{\dagger\dagger}$: $2.9\times 10^{11}$~cm$^{-2}$}\\
\bottomrule
\end{tabular}

\begin{minipage}{0.95\textwidth}
	\footnotesize
	\textbf{Note} -- The quoted uncertainties represent the 16$^{th}$ and 84$^{th}$ percentile ($1\sigma$ for a Gaussian distribution) uncertainties. Upper limits are given as the 97.8$^{th}$ percentile (2$\sigma$) value. Parameters in brackets were held fixed to the 50$^{th}$ percentile value.\\
	$^\dagger$Column density values are highly covariant with the derived source sizes.  The marginalized uncertainties on the column densities are therefore dominated by the largely unconstrained nature of the source sizes, and not by the signal-to-noise of the observations.  See Supplementary Figures~\ref{ulim_corner_cpd}-~\ref{ulim_corner_pyrr} for covariance plots, and \cite{Loomis:2020aa} for a detailed explanation of the methods used to constrain these quantities and derive the uncertainties.\\
	$^{\dagger\dagger}$Uncertainties derived by adding the uncertainties of the individual components in quadrature.
\end{minipage}

\label{ulim_table}

\end{table*}

\clearpage

\section*{Laboratory Spectroscopy}

Supplementary Tables~\ref{1-cpd_comparison} and \ref{2-cpd_comparison} provide comparisons of the spectroscopic constants determined in this work to those from the literature for 1-cyano-CPD and 2-cyano-CPD, respectively.

\begin{table}[hb!]
    \centering
    \caption{Best-fit spectroscopic constants of 1-cyano-CPD, derived using a $A$-reduced Hamiltonian in the $I^r$ representation,  in comparison to earlier studies.   Constants are given in MHz, with values in parentheses corresponding to 1$\sigma$ uncertainties.}
    \begin{tabular}{l l l l}
    \toprule
    Constant & This Work & Ref.~\citenum{ford:326} & Ref.~\citenum{sakaizumi:3903} \\
    \midrule
    $A$                         &   8352.98(2)          & 8356(5)        & 8353.97(9)  \\
    $B$                         &   1904.2514(3)        & 1904.24(2)     & 1904.24(1)   \\
    $C$                         &   1565.3659(3)        & 1565.36(2)     & 1565.38(1)   \\ 
    $10^3 \times D_J$           &      0.0701(15)       &  $\cdots$      &  0.08(3) \\  
    $10^3 \times D_{JK}$        &      2.361(15)        &  $\cdots$      &  2.46(7) \\  
    $10^3 \times d_{J}$         &      0.0120(11)       &  $\cdots$      &  $\cdots$ \\  
    $10^3 \times d_{K}$         &      1.21(13)         &  $\cdots$      &  $\cdots$ \\  
    $\chi_{aa}$(N)              &      -4.1796(21)      &  $\cdots$      &  $\cdots$ \\  
    $\chi_{bb}$(N)              &       2.3052(26)      &  $\cdots$      &  $\cdots$ \\  
    \bottomrule
    \end{tabular}
    \label{1-cpd_comparison}
\end{table}

\begin{table}[hb!]
    \centering
    \caption{Best-fit spectroscopic constants of 2-cyano-CPD, derived using a $A$-reduced Hamiltonian in the $I^r$ representation, in comparison to earlier studies. Constants are given in MHz, with values in parentheses corresponding to 1$\sigma$ uncertainties.}
    \begin{tabular}{l l l l}
    \toprule
    Constant & This Work & Ref.~\citenum{ford:326} & Ref.~\citenum{sakaizumi:3903} \\
    \midrule
    $A$                          &  8235.66(4)          & 8235.0(13)     & 8302.7357 \\
    $B$                          &  1902.0718(2)        & 1902.07(2)     & 1902.3323 \\
    $C$                          &  1559.6502(2)        & 1559.67(2)     & 1562.5153 \\
    $10^3 \times D_J$           &      0.0561(37)       &  $\cdots$      &  0.07(3) \\  
    $10^3 \times D_{JK}$        &      2.286(46)        &  $\cdots$      &  2.459(8) \\  
    $\chi_{aa}$(N)              &      -4.234(6)        &  $\cdots$      &  $\cdots$ \\  
    $\chi_{bb}$(N)              &       2.236(7)        &  $\cdots$      &  $\cdots$ \\ 
    \bottomrule
    \end{tabular}
    \label{2-cpd_comparison}
\end{table}

\end{document}